\documentclass[]{article}

\usepackage[numbers,sort&compress]{natbib}% Citation support using natbib.sty
\usepackage{graphicx}
\bibpunct[, ]{[}{]}{,}{n}{,}{,}% Citation support using natbib.sty
\begin{document}

\title{Classifying Contaminated Cell Cultures Using Time Series Features}

\author{
Laura L. Tupper\thanks{Corresponding author. Mount Holyoke College, 50 College St., South Hadley, MA 01075, ltupper@mtholyoke.edu}, Charles R. Keese\thanks{Applied BioPhysics, Inc., 185 Jordan Road, Troy, NY 12180}, and David S. Matteson \thanks{Cornell University, 1198 Comstock Hall, 129 Garden Ave., Ithaca, NY 14853}
}

\maketitle

\begin{abstract}
We examine the use of time series data, derived from Electric Cell-substrate Impedance Sensing (ECIS), to differentiate between standard mammalian cell cultures and those infected with a mycoplasma organism. With the goal of interpretable results, we perform low-dimensional feature-based classification, extracting application-relevant features from the ECIS time courses. We can achieve very high classification accuracy using only two features, which depend on the cell line under examination. Initial results also show the existence of experimental variation between plates and suggest types of features that may prove more robust to such variation. Our paper is the first to perform a broad examination of ECIS time course features in the context of detecting contamination; to combine different types of features to achieve classification accuracy while preserving interpretability; and to describe and suggest possibilities for ameliorating plate-to-plate variation. \\

\end{abstract}

%%%%%%%%
\section{Introduction}
\label{sec:intro}
%%%%%%%%

%%%%
\subsection{Data challenge: cell contamination}
%%%%

The study of cells in culture is a vital component of biological research, allowing the examination of cells' physical morphology, their patterns of growth and progression through the life cycle, and their responses to stimuli and the environment. Yet there is a reproducibility crisis in cell culture research, in large part due to misidentification and contamination of cell samples (see \cite{hughescontam}). One common issue is the contamination of mammalian cells with other microorganisms such as \emph{Mycoplasma}, which can flourish in the medium used to grow the cells and create misleading results. 

In this paper, we address the issue of contaminated cell cultures as a classification problem, with two goals: to achieve high classification accuracy of cells as infected or uninfected, but also to produce results that are easy to interpret, visualize, and intuit in the context of the application. We draw on an automated, non-invasive data collection method, \emph{electric cell-substrate impedance sensing}, to generate time series corresponding to individual cell cultures. Several typical methods for such high-dimensional data -- such as multivariate logistic regression, PCA, and classification based on a large number of features -- are not appropriate since they are in conflict with the goal of easy interpretability. Instead, we turn to feature-based methods, using characteristic features derived from these time series to classify the corresponding cell cultures.

The structure of the paper is as follows. In the current section, we introduce the data collection method and the specific datasets used in this paper. Section \ref{sec:features} describes our approach: the general methodology of low-dimensional feature-based time series classification, and the types of features we generate from the data, with references to the previous studies that suggested these features. In Section \ref{sec:class} we classify individual cultures as infected or uninfected and examine the performance of different types of features, while Subsection \ref{sec:variations} extends the investigation to different cell types and addresses the possibility of experimental variation. Conclusions and suggestions for future research appear in Section \ref{sec:conclusion}. The online supplementary materials provide more details on the methodology of data collection, a complete list of features investigated, and additional classification performance results.

%%%%
\subsection{Relevant data}
\subsubsection{Electric Cell-substrate Impedance Sensing}
%%%%

Electric cell-substrate impedance sensing, or ECIS{\small{\textregistered}}, is an established tool for measuring the behavior and characteristics of cells over time. An introduction can be found in \cite{giaever91}; since then, the technique has been used in many studies in a variety of biological contexts. \cite{lukic} contain a survey of many of these studies, while \cite{hong} provide an overview of the use of ECIS in cancer studies.

The outline of the technique is as follows. First, an experimenter inoculates a tray of several wells with a cell culture. In the bottom of each well are electrodes, smaller than the entire surface of the well. The tray is placed in a machine that passes a weak AC current through the electrodes, at a chosen frequency, and measures the impedance. With appropriate equipment, the impedance can also be decomposed into its two complex components, resistance and capacitance. As the cells multiply and move, they cover more of the electrodes, and the nature of their attachment to the medium underneath them may also change. Eventually they reach confluence, when there is complete coverage of cells across the entire well. All of these processes change the observed impedance and its components at various frequencies. \cite{deblasio} demonstrate that ECIS-based and optical determinations of cell confluence coincide, a welcome result since ECIS is far less labor intensive than optical assessment of cell cultures, and requires no chemical labels or markers to be applied to the cells.

The ECIS process is non-invasive, and does not damage the cells or change their morphology, allowing for sustained observation of the same culture. If desired, however, the machine can \emph{wound} the culture, by passing a high-voltage current through the electrodes. This process kills most or all of the cells located on top of the electrodes. We can then observe further changes in impedance as the dead cells are replaced by living cells moving in from outside the edges of the electrodes. ECIS observations can be considered as time series; these time series are high-dimensional if we consider each well to have measurements on multiple frequencies linked by time point, or if we consider both components of impedance at each time point.

% fig: sample ECIS data
\begin{figure}[h]
    \centering
    \includegraphics[width=\textwidth]{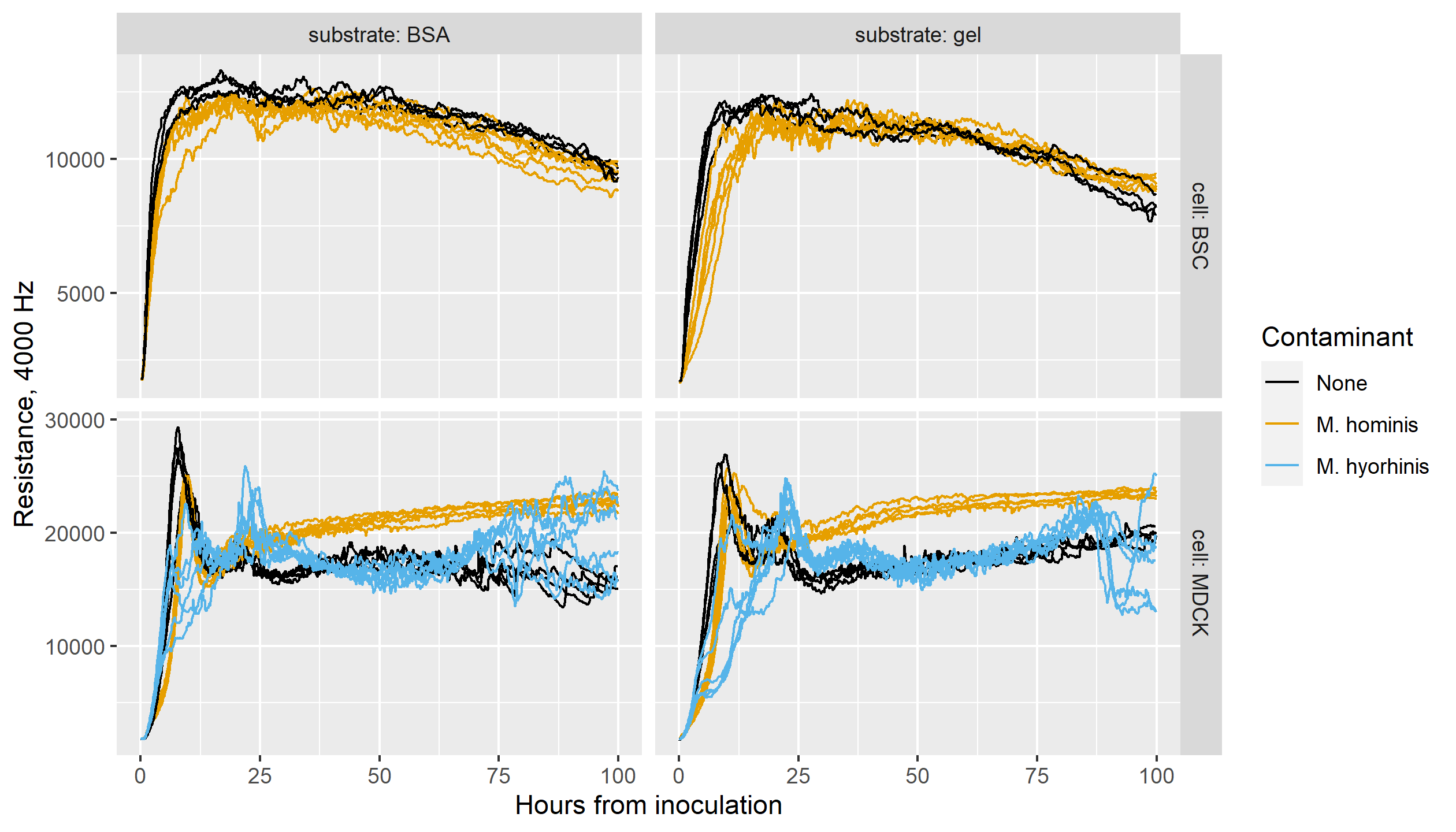}
    \caption{Sample ECIS data: time courses of resistance for different cell types.}
    \label{fig:demoTS}
\end{figure}

Many studies have used ECIS data to distinguish between specified cell types, or to observe differences in a cell before and after it undergoes some process, such as electroporation (\cite{stolwijk}), damage (\cite{heijink}), transformation into cancer cells (\cite{park}), or progression to a new phase of the cell cycle (\cite{wang}). But quantitative discussion of the data has been sparse. \cite{gelsinger} appear to be the first to examine quantitative features of ECIS data across multiple cell lines, with the goal of identifying the cell types in unknown or potentially-mislabeled cultures. \cite{zhang} examine contamination data as we do in this paper, comparing long-memory behavior and the timing of the confluence stage in infected and uninfected cells. We build on this work by bringing in many other types of features, and doing a quantitative comparison of their efficacy in identifying contamination in various cell types; this paper expands on our conference presentation in \cite{tuppermilets}.

\subsubsection{The current dataset}

In this paper, we use data provided by Applied BioPhysics, Inc., generated using the ECIS Z$\theta$ machine. Here we describe the details of the datasets.

Cells are grown on arrays, or plates; each plate contains 96 wells, allowing the observation of multiple cell types and growth conditions within a single plate. At each time point, the ECIS Z$\theta$ equipment generates four measurements: $Z$, impedance; $R$, the resistive component; $C$, the capacitive component; and $H$, the time in hours corresponding to the current time index. These measurements are obtained for each well using several different AC frequencies, from 500 to 32000 Hz.

The hourly time corresponding to a given time index will be slightly different for each well and frequency. The ECIS machine has a limited measurement speed, and cycles through each frequency on a given well before moving on to the next well; so there will be different actual times associated with different wells and frequencies at a given time point. There is, accordingly, a tradeoff between the time resolution of a dataset and how many wells and frequencies we can observe. In the current dataset, which uses only a handful of frequencies, the differences in precise time between observations at the same time point are negligible overall.

The cultures in our dataset are of two types: the target cell line, and the same cell line contaminated with a mycoplasma organism. We will first examine MDCK II cells, some of which are contaminated with \emph{M. hominis}. Cells are grown on one of two substrates: either with a gelatin coating, or with an adsorbed layer of BSA (bovine serum albumin). Some wells of each type are left empty, containing substrate but no cells, to provide a baseline. Finally, some wells of each type are subjected to wounding, allowing us to observe the process of recovery as new cells replace those killed by the high-voltage pulse.

As an extension to our original investigation, we also examine cultures from a different cell line, BSC-1, some of which are infected with \emph{M. hominis}, and MDCK II cells infected with a different species of mycoplasma, \emph{M. hyorhinis}. A description of these observations can be found in Subsection \ref{sec:variations}.

%%%%%%%%
\section{Methodology: feature-based classification}
\label{sec:features}
%%%%%%%%

\subsection{The feature-based approach}

In feature-based classification, instead of calculating similarity or distance ``pointwise" between individual time points in the time series, we reduce the effective dimension by generating a limited set of features from the original data. We may then perform classification or clustering using these features, or a subset of them. There is an extensive body of literature on feature generation and selection; an overview and typography of many approaches appears in \cite{dash}. \cite{maharajdursobook} provide a discussion of feature-based approaches specifically for the classification of time series data. One strength of the feature-based approach is the reduction of dimensionality, as argued in \cite{wanghyndman}, which can avoid many of the problems involved in high-dimensional clustering (for examples, see \cite{aggarwal}). We also avoid the problematic requirement that all the time series under consideration be exactly the same length, as mentioned in \cite{keoghcompression}. 

The feature-based approach is designed to provide more interpretable results, since the user need only consider a few features to understand why an observation is classified in a particular way. In particular, we use \emph{pairs} of features to perform classification. This means that we can easily visualize the distribution of scores across observations, and also visualize the classification regions in a two-dimensional feature space.

In this paper, we generate a variety of features from the ECIS time courses. We have attempted to designate features that reflect characteristics of ECIS data mentioned as useful in previous studies -- though these previous studies did not attempt to combine different types of features. Many of these are simple characteristics of the time series, such as the level at a particular time. We have also included some features based on time series analysis and those aimed at capturing particular stages of the cells' behavior. All computation was performed using R (\cite{Rgeneral}) with the \texttt{tidyverse} collection of packages (\cite{Rtidyverse}); estimation details for individual features are given in the supplementary materials.

\subsection{Deriving features from ECIS data}

While the shape of the ECIS time series depends on the cell line, the substrate on which it is grown, and any contamination, there are general patterns that inform our choice of features. Immediately following the initial inoculation of the well, there is an attachment phase as the cells attach themselves to the substrate, then a growth period as cells reproduce and move to cover the electrode. These correspond to increasing resistance values in the data. Once the cells reach confluence -- possibly following a drop in resistance from an initial peak -- large-scale movement and spreading becomes impossible, and we see a long period of more consistent values. During this phase, the cells exhibit \emph{micromotion}, small movements that cause fluctuations in the resistance, and may also show a gradual trend. After considerable time, usually beyond 72 hours, the cells may begin to senesce, leading to more substantial changes in resistance. The exact time at which the cells reach confluence, and the direction of the overall trend in resistance during confluence, depend on the cell type. Across cells, however, the period between 24 and 36 hours can be safely considered a "confluent window," during which the time series may be treated as stationary after first differencing to remove the trend.

Many analyses of ECIS data in the literature simply discuss the overall level of impedance (or one of its components) as contrasted between cell types. Accordingly, we include several features that reflect the level of resistance at a particular time, measured separately at each frequency.

\cite{park} compare cell lines with their transformed (cancerous) versions, and observe that resistance of the cancerous cells ``increased more rapidly" than that of their noncancerous counterparts. A similar feature appears in \cite{kowolenko}: they discuss the ``initial increase in resistance" as distinguishing between control cells and cells that are treated with compounds encouraging attachment. 

We define two features to reflect the rate of increase of resistance early on. First, we simply record the resistance level at 2 hours after inoculation; this time is during the early attachment and spreading stage, and the resistance at this point gives a measurement of how quickly the cells are spreading. We then calculate the difference between the levels at 7 hours and 2 hours after inoculation, which effectively measures the slope during the growth period but after the very early attachment phase. It is also evident that the early growth in resistance is not linear. Accordingly, we fit a quadratic to the first 2 hours of values after inoculation, and extract the first- and second-degree coefficients from the fit. All these early measurements may be particularly useful in contexts where results must be obtained quickly, as in \cite{rutten}, where practitioners have only a few hours to determine whether stem cells will be functional for a particular recipient before the cells become non-viable.

Moving past the early growth stage, the confluence level of resistance is used in \cite{kowolenko} and \cite{heijink}, while a late-stage measurement of capacitance appears in \cite{baganinchi}. We measure the resistance level at 24 hours after inoculation, which serves as an estimate of the level at confluence. \cite{park} also note that cancerous cells ``peaked higher" than noncancerous cells. We adopt this feature by calculating the maximum level of resistance over the first 24 hours, and over the first 48 hours. Using a uniform time window is important, since some types of cells have upward trends in resistance during confluence and may reach their maximum value at the end of the observed window. As an alternative that does \emph{not} reflect such behavior, we also measure the time and value of the \emph{first} peak following inoculation in a smoothed version of the time series.

To characterize the post-confluence behavior more generally, we turn to time series methods, focusing on the stable confluent period between 24 and 36 hours. Over this range, we fit two different ARIMA models, $ARIMA(1,1,0)$ and $ARIMA(0,1,1)$, and extract their fitted coefficients. Both these models contain a nonzero $d$ or differencing component to account for any overall trend. The estimates of error variance from these models can also be useful in reflecting short-term erratic behavior, as introduced by \cite{tongdabas} and described below. Using model coefficients and errors for clustering has appeared in a variety of time series application areas (for example, \cite{maharajAR} and \cite{dursogiovanni}) and is discussed more fully in \cite{maharajdursobook}. We also take an intermediate approach, analyzing the time series' short-term memory without fully fitting a model, by examining the values of the ACF and PACF at small lags for the differenced series. An ACF-based distance measure for time series clustering is described fully in \cite{galeanoACF}, and related examples and extensions can be found in \cite{wanghyndman}, \cite{caiadopgram} (introducing the PACF as well as the ACF), and \cite{dursoACF} (extending the method to fuzzy clustering). Though these methods have been used in other time series areas, they appear to be new to the specific context of ECIS.

Returning to the biophysics literature, several studies address cells' responses to wounding. Some refer to ``the ability of cells \dots to close an experimental lesion," as \cite{lukic} put it. In the same vein, \cite{heijink} observe resistance levels when confluence is regained after cells are damaged with cigarette smoke. \cite{stolwijk} provide a similar example of examining cells' recovery from an event, though in this case the event is non-fatal electroporation. In this study, the cells' full recovery is mentioned, as well as variation in time to recovery ``dependent on cell type and age of the culture." Time to recovery also appears in \cite{heijink}, as does the reduction in resistance at one to two hours after wounding.

In our dataset, wounding is performed at approximately 48 hours after inoculation, and we define several features to reflect post-wounding behavior. First, we record the minimum level of resistance at (or just following) wounding, since the residual resistance from dead cells can provide information about their type or morphology. To represent the rate of recovery after wounding, we use the difference between the levels at 57 and 52 hours after inoculation; we also take the ratio of these quantities. In addition, we take the ratio of this post-wounding slope to the slope during the pre-wounding growth stage (the difference in levels at 7 and 2 hours after inoculation). 

At high frequencies, we observe a particular pattern in the post-wounding resistance levels. After the sharp drop in resistance in response to wounding, there is a period of swift growth to a local peak, followed by a drop to a more moderate level, and finally a long period when the level is stable or slightly increasing. Although the final resistance levels of infected and uninfected cells may be similar, the characteristics of this first post-wounding peak are different. Accordingly, we include features corresponding to both the value of this peak and the time when it occurs.

Several early papers by Giaever and others, such as \cite{giaever91}, examine cells' micromotion behavior, which can be observed in ECIS data as small fluctuations after a stable level has been reached. \cite{lukic} mention examining micromotion on a 30-minute window, which helps account for any lingering overall trends in level. There appears to be no consensus among researchers on the best way to quantify micromotion. We measure small-scale variation during confluence by looking at the variance and standard deviation of each time course during a post-confluence time window (24 to 36 hours after inoculation), after differencing the values to remove trend. For comparison, we also include the estimated error variance after fitting the ARIMA models mentioned above.

Some papers have noted potential long-memory behavior in ECIS time series. The DFA (detrended fluctuation analysis) approach to quantifying micromotion appears in \cite{lovelady} in examining cells' response to a toxin, and in \cite{schneider} when comparing cells before and after a phenotypic transition. Recently, \cite{zhang} have performed a more involved analysis of the regime change between early growth and confluence. Their \emph{Growth-to-Confluence Detector} assumes that early growth follows a nonlinear growth curve with heteroscedastic noise, while the confluence stage follows a long memory model. Following their methodology, we extract the estimated long memory parameter from post-confluence data, as well as the estimated time of the change point from growth to confluence behavior.

\subsection{Variations of feature types}

For the broad feature search, we calculate each feature on the time course of resistance. Basic feature types (maxima, values at a set time, and features combining these) are found both for the raw time courses and for a smoothed version, using a rolling windowed average over 5 consecutive time points. This windowing is intended to make the feature scores robust to momentary noise in the data.

We also introduce a normalization procedure to help account for variations in conditions over time. Each tray contains several ``empty" wells, which are treated with a protein coat (BSA or gel) but not inoculated with cells. The time courses for these wells can be considered as a baseline, reflecting the resistance created by the electrode and substrate under the current laboratory conditions. For each type of protein coat, we average the time courses across all empty wells on the plate, and take the ratio of each well's time course to this baseline. After calculating many features both on the original time courses and on these normalized values, we have found that this normalization has only minimal impact on the feature scores; thus some more advanced features are calculated only using the original data.

All features are calculated for all frequencies available in the dataset, on the grounds that measurements at different frequencies may reflect different components of cell behavior. This idea is discussed in \cite{lukic}, and some studies make reference to particular frequencies of interest; for example, \cite{wang} find 60 kHz to be most useful in their study of the cell cycle.

\subsection{Classification process}

We attempt three methods for performing classification on the feature data: classification trees, linear discriminant analysis (LDA), and quadratic discriminant analysis (QDA). A thorough discussion of these methods appears in \cite{gelsinger}, including an exploration of different parameter choices for discriminant analysis -- which is found not to have a substantial impact on the results. In our own analysis, we observe that LDA generally performs the best of the three methods, or close to it; and in general LDA may be preferred for these fairly small samples since there are fewer parameters to estimate. Accordingly we present detailed results based on LDA. Further notes on classification methods can be found in the supplemental materials.

We perform classification using each possible pair of features, in turn; from the 476 features and feature variations, we obtain over 113,000 possible feature pairs. For each feature pair, we split the observations evenly into training and testing sets. We train an LDA classifier using the training observations, use it to classify the testing observations as infected or uninfected, and record the success rate of these classifications. We then repeat the process with a different training/testing split, a total of 10 times. The performance values discussed below use the average over all 10 splits.

%%%%%%%%
\section{Application and results}
\label{sec:class}
%%%%%%%%

\subsection{General classification results}

\begin{figure}[h]
    \centering
    \includegraphics[width=0.6\textwidth]{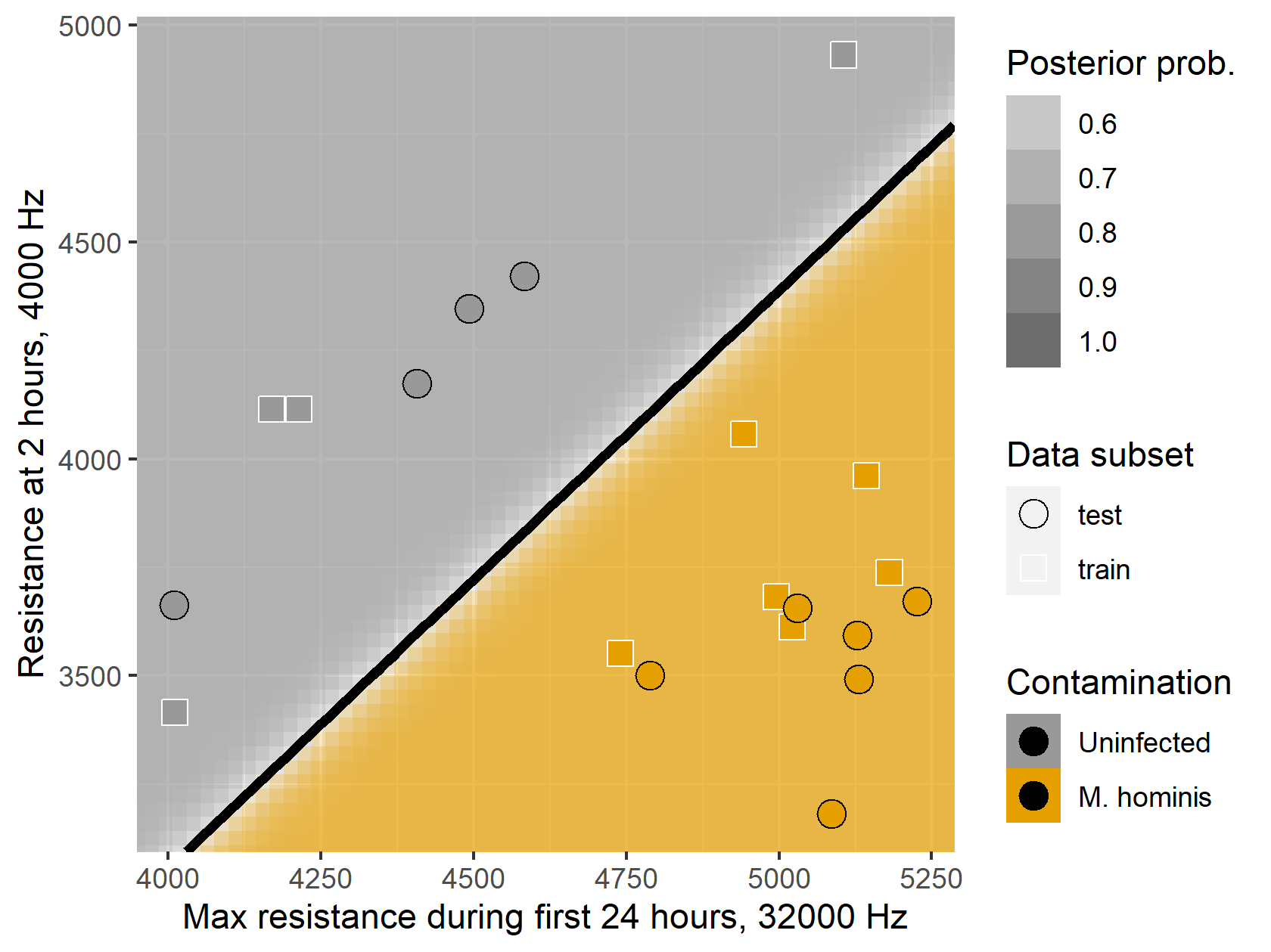}
    \caption{LDA classification regions based on one pair of features. This pair yields perfect classification of the test observations.}
    \label{fig:region_LDA_MhomMDCK_63_4}
\end{figure}

An example of the LDA classification results is shown in Figure \ref{fig:region_LDA_MhomMDCK_63_4}, using two simple features: resistance at 2 hours after inoculation (measured at a frequency of 4000 Hz) and the maximum resistance during the first 24 hours after inoculation (at 32000 Hz). The classifier uses the training observations (square points) to generate a posterior probability of contamination for each point in the feature space, and a linear division of the feature space into infected and uninfected regions. The test observations (round points) are all correctly classified based on their coordinates in this feature space, and we can see that there is very clear separation between contaminated and uncontaminated cultures in terms of these feature scores.

% demo class region

\begin{figure}[h]
    \centering
    \includegraphics[width=0.65\textwidth]{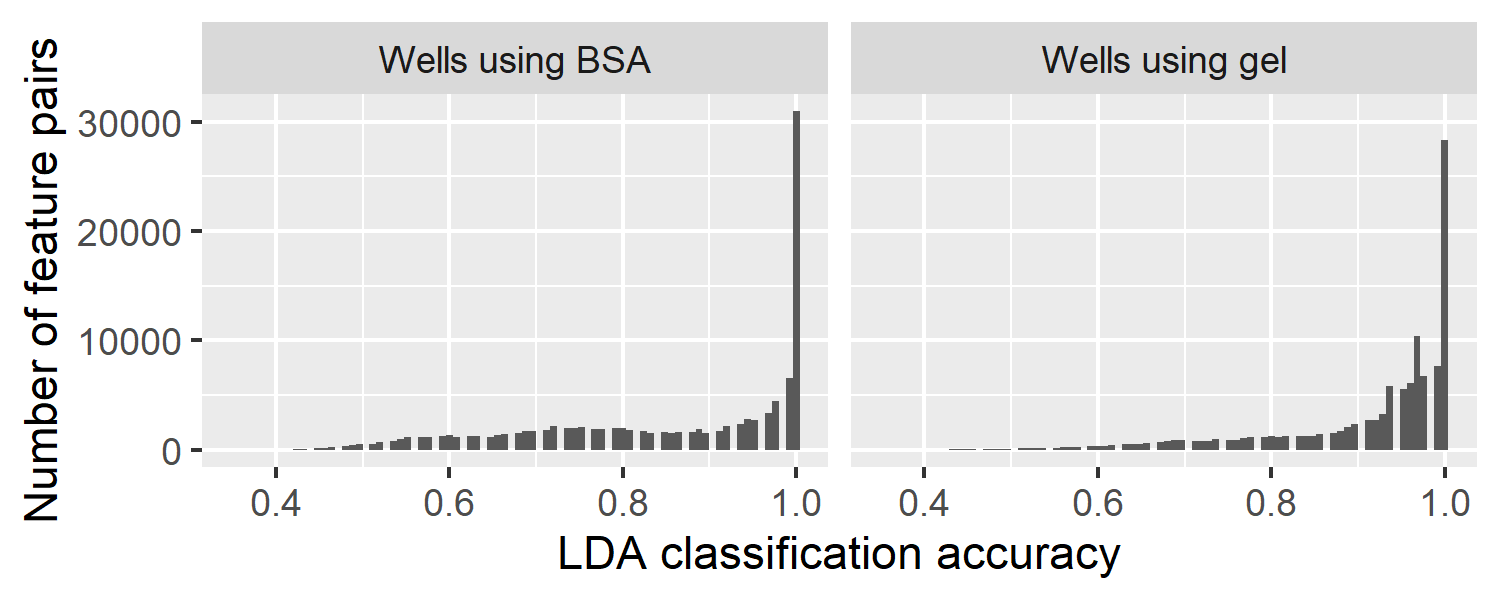}
    \caption{Histogram of classification accuracy values for each pair of features. Many feature pairs yield perfect accuracy.}
    \label{fig:acc_hist_MhomMDCK_LDA}
\end{figure}

In general, classification accuracy is high. There is no ``best pair" of features: many different feature pairs offer strong, and in many cases equal, performance. For example, on our sample of BSA-treated wells, about 27\% of all feature pairs yield perfect classification accuracy. The proportion of perfect-accuracy pairs is about 25\% on gel-treated wells, though there are more feature pairs that yield \emph{near-}perfect accuracy on gel, as shown in Figure \ref{fig:acc_hist_MhomMDCK_LDA}. 

% hist of pair accuracy values

The protein coat on which the cells are grown does appear to affect their behavior, and in particular, the differences in behavior between infected and uninfected cells: some features are more effective for classification of cells grown with BSA, while others are more effective on gel-treated wells. Since it is a simple matter in practice to grow cells using one or the other substrate, high performance on either group of cultures is equally useful; when giving performance values, we will indicate on which substrate the performance is achieved.

This is a promising indication of the effectiveness of ECIS measurements for distinguishing contaminated cultures, using simple features in a low-dimensional space that is easy to visualize. It is worth noting, though, that the sample size is small, and the problem becomes more difficult once we introduce the possibility of experimental variation across plates, as discussed in Subsection \ref{sec:variations}.

\subsection{Feature combinations}

% heatmap of some pair accuracy values
\begin{figure}[h]
    \centering
    \includegraphics[width=0.95\textwidth]{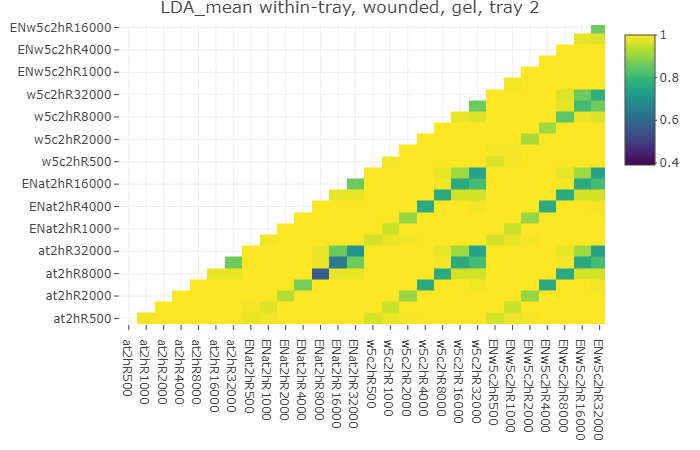}
    \caption{Heatmap of classification accuracy (on gel) for a few feature pairs, showing ``sub-diagonals" of feature pairs with poorer performance.}
    \label{fig:acc_heat_subdiag}
\end{figure}

Figure \ref{fig:acc_heat_subdiag} shows a heatmap of performance values for a subset of feature pairs, with the brighter tiles representing feature pairs with higher accuracy. This subset illustrates the high performance that can be achieved with many different features, but also suggests rules for combining features, based on the ``sub-diagonals" of poorer performance that can be seen throughout. 

Some of these low-performance areas correspond to pairs consisting of a feature and a modified version of the same feature: for example, \emph{at2hR32000} (resistance at 2 hours, 32000 Hz) paired with \emph{w5c2hR32000} (resistance averaged over a window at 2 hours, 32000 Hz). These feature modifications do not greatly change the information in the feature, so that pairing a feature and its modified version is essentially equivalent to using a single feature for classification. This behavior suggests that windowing is optional, though it may still be desirable for robustness against recording errors. Other low-performance cases appear when the pairing consists of the same feature calculated on two different versions of the data: the original time series and the version normalized to empty wells (features beginning with ``EN"). Again, this shows that normalizing the time series has only a minimal effect on the features calculated from it.

To a lesser extent, we also see relatively weak performance when a feature is paired with itself at a similar frequency, or with a modification of the same feature at a similar frequency. For example, \emph{at2hR32000} and \emph{at2hR16000} are not an effective pair, but each yields perfect classification accuracy if paired with \emph{at2hR500}. This provides further evidence that more divergent frequencies provide more useful information, by reflecting different characteristics of the cells' behavior.

\subsection{Individual feature performance}

We can obtain a simple measure of the effectiveness of individual features by averaging the classification accuracy over all pairs involving a given feature. A few notable points of performance are described below.

\emph{Early behavior.} High classification accuracy can be achieved with features that use only the first several hours of data following inoculation. The fitted quadratic coefficients for the first two hours of the time series have excellent average accuracy for low frequencies (averaging above 0.99 on gel for frequencies up to 2000 Hz), while a simple difference between the resistance levels at 7 hours and 2 hours after inoculation is also effective (average accuracy 0.99 on gel at 32000 Hz, though several other frequencies work nearly as well). Even the level at two hours, by itself, performs well in this case (for example, 0.99 on BSA at 1000 Hz), though this early level can be sensitive to variation in the environmental conditions at inoculation.

\emph{Peak values.} Information about the peak values of resistance can be effective in classification, particularly the peak reached after recovering from wounding. The value of this post-wound peak gives an average accuracy above 0.99 on gel for all frequencies up to 2000 Hz. The value of the first peak after inoculation can also be useful (0.99 on gel at 2000 Hz). Identifying the first peak is slightly more effective than using the maximum value over a fixed time period (the best performance of this type is 0.98 on gel at 2000 Hz). In many cases these two features will identify the same peak, but for cell types with a post-inoculation peak and drop followed by an upward trend, using the maximum over a fixed window may inadvertently reflect the later post-confluence behavior rather than the initial peak.

\emph{Time series features.} The coefficients from ARIMA models (fitted to the post-confluence period when the time series can be considered integrated of order 1) do not perform well in general. The cell cultures do display some short-memory behavior, and their fitted ARIMA models are substantially different from white noise and from models fitted to empty wells; but infected and uninfected cells do not have reliably distinct model coefficients. (Aside from the constant coefficients, which are arguably overly-complicated ways of describing the trend, the best performer is the $\hat{\theta}$ coefficient from an ARIMA(0,1,1) model at 32000 Hz, with an average accuracy of 0.94 on gel.) The values of the ACF and PACF seem to be a more effective way of characterizing the short-memory behavior; for example, the second lag of the PACF at 32000 Hz has an accuracy of 0.97 on gel. Fitting the ARIMA models, however, does allow an alternative means of estimating local noise as a measurement of cell micromotion after confluence.

\emph{Post-confluence behavior.} We measure cells' post-confluence behavior in two main ways: micromotion as reflected in the local variability of the time series, and long-memory behavior. Both seem to be an effective basis for classification. We obtain average accuracy between 0.97 and 0.98 on gel with any of three measures of local variability (the sample variance of the differenced series; the estimated $\hat{\sigma}^2$ from an ARIMA(1,1,0) model; and the estimated $\hat{\sigma}^2$ from an ARIMA(0,1,1) model). Notably, all three feature types perform best at the highest frequency, 32000 Hz. The estimated long-memory parameter for the post-confluence time series, meanwhile, gives an accuracy of 0.98 on gel at 2000 Hz.

\emph{Post-wounding recovery.} Features describing the cells' recovery from a wounding pulse can be very effective. For example, the level at 57 hours minus that at 52 hours represents the rate of recovery shortly after wounding; this has an accuracy of 0.99 on BSA at 500 Hz. It is also effective to compare this post-wounding recovery rate to the initial growth rate: the \emph{ratio} of (57 - 52 hours) to (7 - 2 hours) has an average accuracy of 0.99 on gel for all frequencies up to 4000 Hz. 

%%%%%%%%
\subsection{Extensions of the problem}
\label{sec:variations}
%%%%%%%%

\subsubsection{Cell and contaminant types}

We extend our analysis by examining cultures taken from a different cell line, as well as those infected with a different species of mycoplasma. For reasons of space, a detailed description is not included here; we focus on comparison with the original results, which used MDCK II cells with \emph{M. hominis}. For a cell comparator, we use BSC-1 cells. These cells have a different growth and confluence pattern than MDCK II cells, notably without the large peak prior to confluence. As before, we infect some of these cells with \emph{M. hominis}. We also examine MDCK II cells infected with another type of mycoplasma, \emph{M. hyorhinis}. It is immediately evident that the classification problem depends on both the type of cell and the species of mycoplasma; features that perform well for one cell-contaminant combination may not do so for others. 

To distinguish MDCK II cells from those infected with \emph{M. hyorhinis}, the strongest-performing individual features are those focusing on the early peak behavior. In this case, however, it is more effective to use the maximum level in the first 24 or 48 hours after inoculation, rather than attempting to find the first peak; this seems to be because the early growth period is noisier, leading to false local maxima before the real post-inoculation peak occurs. The other notably effective feature type here is the difference between the level at 57 and 52 hours after inoculation, a representation of the rate at which the cells recover from wounding. Since this also appeared as an effective feature for detecting \emph{M. hominis} infection, wounding cells and monitoring their recovery may be a particularly useful approach when it is not certain what kinds of contaminants may be present.

\begin{figure}[h]
    \centering
    \includegraphics[width=0.65\textwidth]{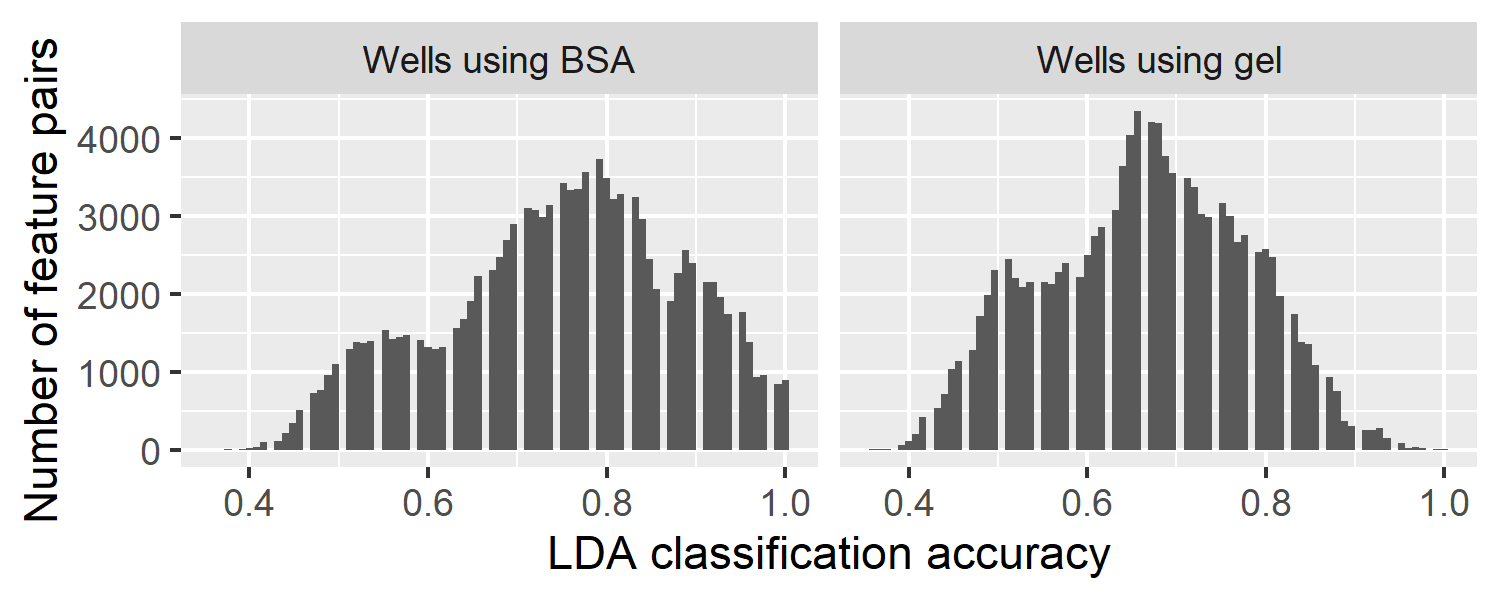}
    \caption{Histogram of classification accuracy values for each pair of features on BSC cells.}
    \label{fig:acc_hist_MhomBSC_LDA}
\end{figure}

For the BSC-1 cells the problem is overall more difficult: as shown in Figure \ref{fig:acc_hist_MhomBSC_LDA}, far fewer feature pairs are able to give good classification performance. Also in contrast to the MDCK II dataset, classification of BSC-1 cells appears to be easier for BSA-treated cultures than for those grown on gel. Due to the growth patterns of BSC-1 cells, the early levels and first-peak features that worked so well for MDCK II cells are not distinctive here. Instead, all of the features with high average performance for BSC-1 cells (above 0.91 on either protein coat) are of two types: coefficients from a quadratic fit to the first two hours of growth, or measurements of local variation/micromotion after confluence. These results underline the importance of establishing the most effective and characteristic features for individual cell lines, and of exploring a variety of possible feature types.

% histogram of accuracy for each pair for MhomBSC

\subsubsection{Experimental variation}

In laboratory use, the cell cultures that require classification may often be grown on a separate plate than the training set of cultures known to be infected or uninfected. The test set may even be measured at a substantially different time, or under different conditions. It is therefore desirable to examine the robustness of classification to variation across plates, and experimental variation in general.

\begin{figure}[h]
    \centering
    \includegraphics[width=0.6\textwidth]{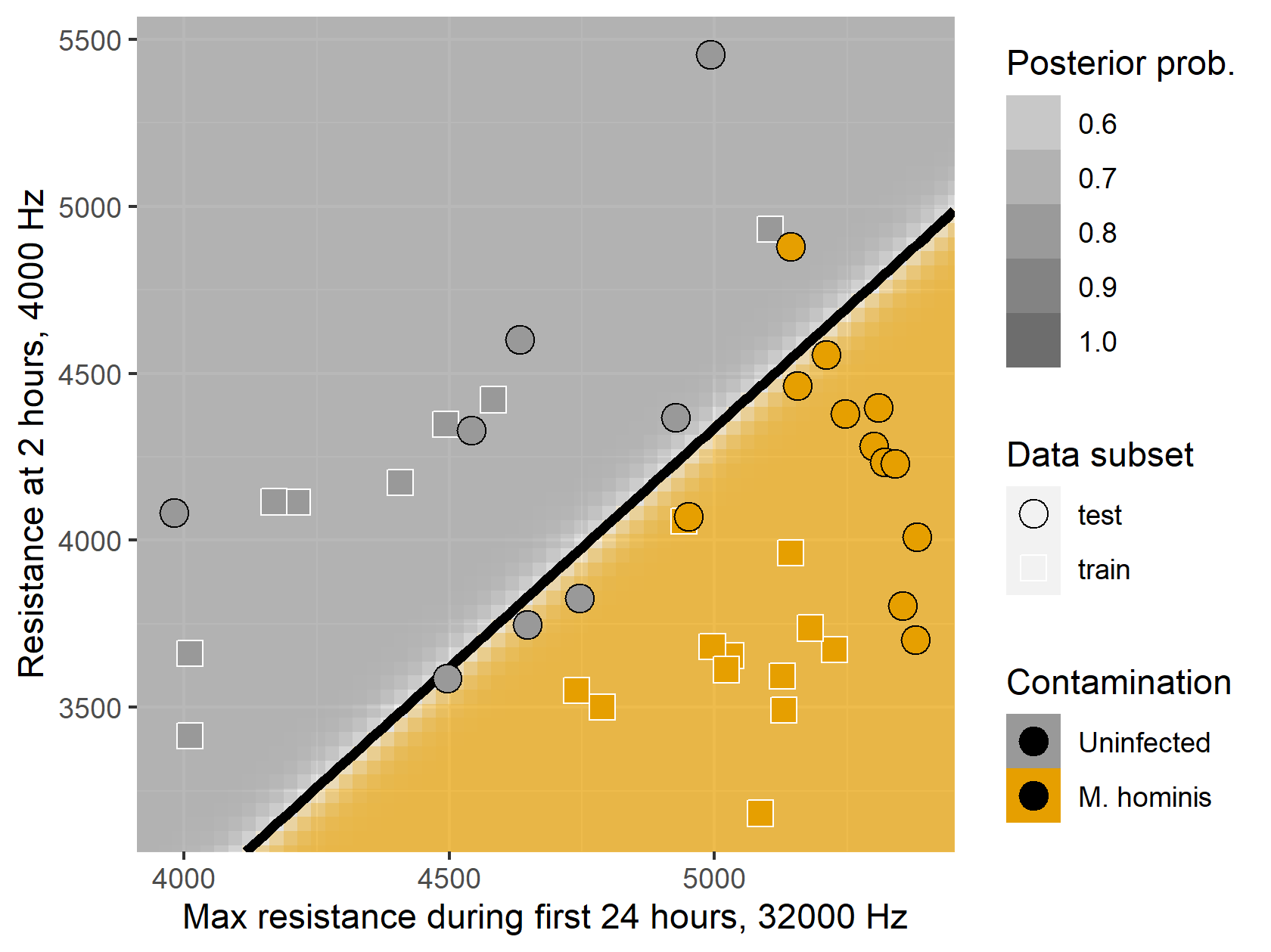}
    \caption{LDA classification regions, with training set from one plate and test set from a second plate.}
    \label{fig:region_LDA_MhomMDCK_63_4_CT}
\end{figure}

To this end, we examine four separate plates of MDCK II cells, each plate containing some uninfected and some infected cultures (using \emph{M. hominis}) taken from the same frozen source, but grown and measured at a separate time. We can immediately see that plate-to-plate variation exists, and makes the classification problem substantially more difficult. Figure \ref{fig:region_LDA_MhomMDCK_63_4_CT} shows the same set of features and classification procedure as Figure \ref{fig:region_LDA_MhomMDCK_63_4}, but here, all observations on one plate are used as the training set, while the test observations come from a different plate. The LDA classifier was able to achieve perfect classification accuracy within a single plate, but cannot do so when working with two separate plates.

% demo class region 

% histogram of accuracy for each pair -- this is averaged over all train-test tray pairs using all 4 trays
\begin{figure}[h]
    \centering
    \includegraphics[width=0.65\textwidth]{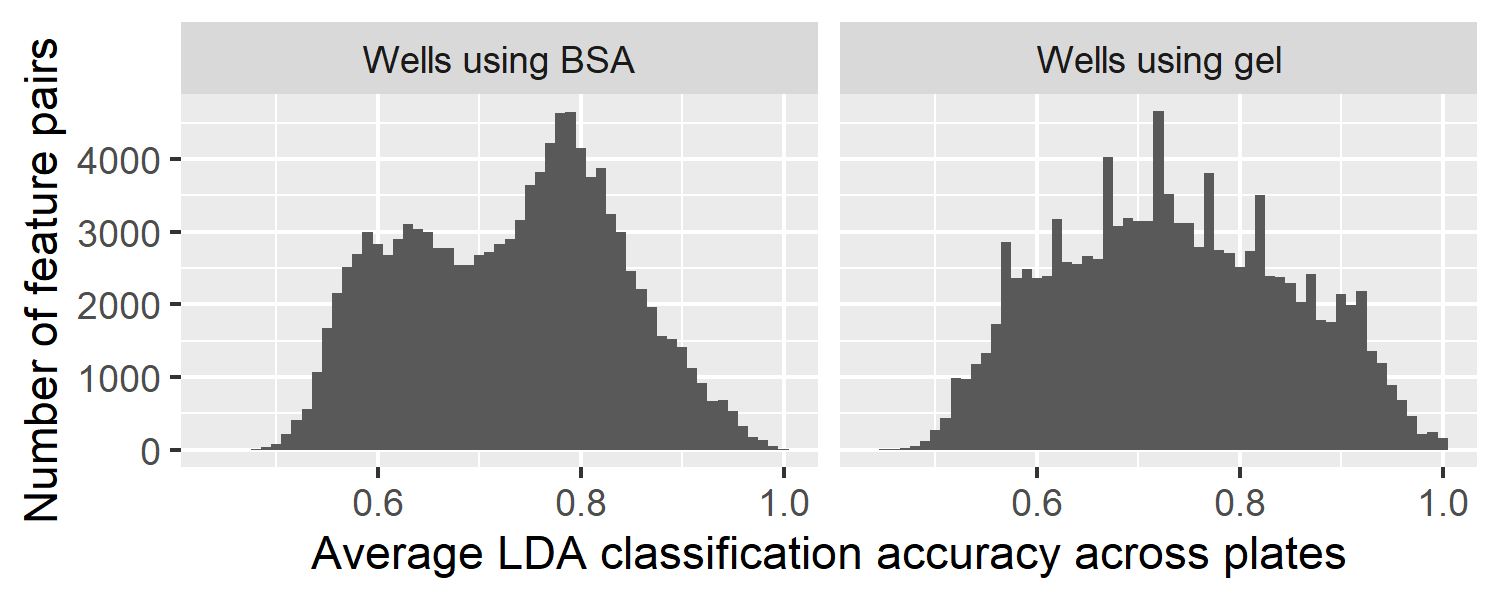}
    \caption{Histogram of classification accuracy values for each pair of features, with training and test observations from different plates. Performance is generally much worse than for within-plate classification.}
    \label{fig:acc_hist_MhomMDCK_LDA_CT}
\end{figure}

In the cross-plate classification problem, the accuracy rates are markedly lower than for the single-plate analyses above. Some results are consistent with those discussed above: ARIMA coefficients still do not perform well, and pairing a feature with its own modified version (at the same frequency) is ineffective.

A noticeable standout is the ``7 hours minus 2 hours" family of features, which reflect the rate of the initial cell growth after inoculation. Though simple, this feature type performs well when paired with several other types of features, or even with itself at two distinct frequencies; for example, it yields an average accuracy of 0.93 on gel-treated cultures at 4000 Hz. Such a feature draws on the information provided by the early growth stage while accounting for the level reached during the first two hours, making the feature more robust to the initial conditions of each cell culture. This may provide greater robustness to experimental differences such as the initial number of cells at inoculation or the starting temperature of the cells.

In general, features focusing on later-stage behavior also appear to be more robust to plate-to-plate variation. The process of each culture reaching confluence, or being wounded with the same high-voltage current, appears to equalize the effects of initial conditions. Indeed, features measuring local variability after confluence are among the top overall performers (such as $\hat{\sigma}^2$ from an ARIMA(0,1,1) model, with an accuracy of 0.92 on BSA at 500 Hz). These are promising indicators of useful features, though more data will be needed to characterize plate-to-plate variation more fully and determine which features are most robust to it.

%%%%%%%%
\section{Conclusion and discussion}
\label{sec:conclusion}
%%%%%%%%

This paper demonstrates a methodology for differentiating between standard cell cultures and those infected with a mycoplasma organism, by comparing features of their ECIS time courses. Detecting contamination in this way is particularly desirable since the ECIS data gathered during the process can also be used to investigate other scientific questions about cell morphology, without requiring a separate tool, or labeling or destroying cells.

We have determined that high classification accuracy can be achieved using a straightforward classification algorithm such as LDA, applied to a small number of interpretable features. Not only can the features themselves be readily illustrated, but using only two features allows for visualization of the feature space and the classification results.

In the basic problem, we achieve very high accuracy rates using any of a variety of feature pairs. Most often, effective pairs combine features measured at different frequencies, which reflect different characteristics of the cells. Several different types of features can be of use, including those reflecting initial growth rates, peak resistance levels, post-confluence micromotion, recovery from wounding, and long-memory behavior. We have also seen, however, that the classification efficacy of specific features depends on the type of cell and the type of contaminant under investigation.

Based on our current dataset, we can also see that there is substantial variation across plates. We have already shown that certain features appear to be more robust to this ``plate effect" than others. Especially promising are features that reduce the effect of initial conditions -- either by ignoring initial impedance levels, by waiting for the confluence stage to be achieved, or by wounding all the cultures and observing their recovery. Since experimental variation is to be expected in practice, where cultures will be grown at different times and in multiple laboratories, this is an important topic for further investigation.

While few features show clear separation between all infected cells (across multiple plates) and all uninfected cells, we see that some features show separation within each tray and a consistent direction of difference across trays. If it is possible to normalize the data for each tray, such features would be effective for cross-tray classification. In a situation where several suspect sources of cells are being compared at once, this could be done by inoculating some wells on each tray with ``baseline" cells from a single source already known to be uninfected. But where experiments are being performed over a long time, or baseline cells are not locally available, some other methodology would be necessary to standardize the trays. In future, we intend to acquire a larger dataset in order to be able to study this tray effect more fully.

%%%%%%%%
\section*{Acknowledgements and disclosures}
%%%%%%%%

The authors gratefully acknowledge financial support from the Cornell University Institute of Biotechnology, the New York State Foundation of Science, Technology and Innovation (NYSTAR), a Xerox PARC Faculty Research Award, National Science Foundation Awards 1455172, 1934985, 1940124, and 1940276, USAID, and Cornell University Atkinson Center for a Sustainable Future.

Charles R. Keese is Chief Scientific Officer of Applied BioPhysics, Inc., which manufactures the ECIS Z$\theta$ machine used to collect the data described in this paper.

%%%%%%%%

\bibliographystyle{plain}
\bibliography{abpcontambib}

\end{document}

\title{Classifying Contaminated Cell Cultures Using Time Series Features: \\ Supplemental Materials}
\author{Laura L. Tupper, Charles R. Keese, and David S. Matteson}

\maketitle

\section{Methodology of ECIS data collection}

\subsection{Preparation of the wells}

Proteins were adsorbed to the surface of wells in 96W1E+ (Applied BioPhysics, Inc.) arrays using sterile solutions of bovine serum albumin (BSA) (Pentex{\small{\textregistered}} brand Miles Laboratories, Inc.) and of gelatin (Fisher purified grade gelatin; 275 bloom) in 0.15 M NaCl at a concentration of 200 micrograms/ml. Half of the wells in each plate received 200 microliters of either BSA or gelatin solutions at room temperature. Following 15 minutes to allow protein adsorption, the solutions were aspirated from the wells, and all wells received 200 microliters of a sterile solution of 10 mM cysteine in distilled water (Electrode Stabilizing solution, Applied BioPhysics) to stabilize the gold electrodes. The cysteine solution remained in the wells for at least 30 minutes and was aspirated just prior to cell inoculation.

\subsection{Cell culture}

BSC-1 cells and MDCK II cells were obtained from the American Type Culture Collection (Manassas, VA) and grown in DMEM (low glucose, SIGMA D6046) with 10\% fetal bovine serum. Antibiotics were not used to eliminate any possible effects upon the mycoplasma infections being studied. Cells were grown to near confluency, and cell suspensions were prepared. BSCI-1 cell layers were disrupted using trypsin/EDTA, and MDCK II were exposed first to EDTA for 10 min before the introduction of trypsin/EDTA. Detached cells were harvested using complete medium and centrifuged (250 x g, 5 mins). The cell pellets were resuspended in complete medium at 37$^\circ$ C, counted, and plated into the ECIS wells. The arrays were mounted in a 96 well ECIS Station that was maintained in a tissue culture incubator at 37$^\circ$ C with 5\% carbon dioxide and 95\% relative humidity.

\subsection{Infection of cell layers with mycoplasma}

Freeze-dried samples of \emph{M. hominis} and \emph{M. hyorhinis} were obtained from the American Type Culture Collection (Manassas, VA) in sealed vials and opened under aseptic conditions. The dried samples were divided into two approximately equal parts.  One was placed in a sealed glass vial and stored under liquid nitrogen vapor for future inoculation.  The other was split and used to inoculate cultures of either BSC-1 cells or MDCK II cells. The lyophilized material was dispersed into the medium of ~ 50\% confluent culture to initiate the infection. All infected cultures were passaged several times using fastidious care to prevent contamination of any uninfected cultures. 

Tests were routinely run to verify infection of the inoculated cultures and to assure that uninfected cultures were mycoplasma free. We used one of two different commercially available kits to detect mycoplasma: the MycoAlert{\small{\textregistered}} Detection Kit by Lonza and the PlasmoTest{\small{\texttrademark}} by InvivoGen.

\subsection{Measurements}

All impedance measurements were obtained using the ECIS{\small{\textregistered}} Z$\theta$ instrumentation using ninety-six well 96W1E+ arrays, provided by Applied BioPhysics, Inc., Troy, NY.  (\cite{giaever84}, \cite{giaever91}, \cite{stolwijk})

Following the addition of cell suspensions to the wells, time course impedance data were gathered at low non-invasive current (2.5 microamperes) at seven different AC frequencies, doubling each frequency from 500 Hz to 32 kHz. The impedance (both the resistive and reactive components) recorded the attachment and spreading of the cells and the formation of a confluent cell monolayer. At 48 hours post-inoculation, half of the wells received a brief invasive high current pulse (3000 microamperes at 60 kHz for 20 sec), killing the cells attached to the electrodes. The system then reverted to the low current mode, and cellular migration was then monitored as the normal cells on the periphery of the electrode migrated inward to reestablish a monolayer upon the electrodes (\cite{keese04}).

\section{List of features}

The following features were extracted from the time series of resistance values for each cell culture. Each feature was calculated using the time series at each available AC frequency. In addition, we calculated most features using both the original measured values, and values normalized to the empty wells, though typically there were only small differences in performance (see "Application and Results" in the main paper).

\begin{enumerate}
    \item Value at 2 hours after inoculation
    \begin{enumerate}
        \item \emph{Windowed} value at 2 hours: average value over 5 consecutive time points, centered at 2 hours after inoculation
    \end{enumerate}
    \item Value at 24 hours after inoculation
    \begin{enumerate}
        \item Windowed value at 24 hours
    \end{enumerate}
    \item Second-order coefficient of a quadratic fit to the first 2 hours after inoculation
    \item First-order coefficient of a quadratic fit to the first 2 hours after inoculation
    \item Maximum value up to 24 hours after inoculation
    \begin{enumerate}
        \item Maximum value up to 24 hours after inoculation of a \emph{smoothed} time series (rolling average over 5 time points)
    \end{enumerate}
    \item Maximum value up to 48 hours after inoculation
    \begin{enumerate}
        \item Maximum value up to 48 hours after inoculation of a smoothed time series
    \end{enumerate}
    \item Features of the first peak in a smoothed time series after inoculation (these features use a rolling window of 15 time points, for reduced sensitivity to noise in the early series):
    \begin{enumerate}
        \item Value at this peak
        \item Time when this peak occurs
    \end{enumerate}
    \item Value at 7 hours minus value at 2 hours
    \begin{enumerate}
        \item Windowed value at 7 hours minus windowed value at 2 hours
    \end{enumerate}
    \item Value at 57 hours minus value at 52 hours
    \begin{enumerate}
        \item Windowed value at 57 hours minus windowed value at 52 hours
    \end{enumerate}
    \item Value at 57 hours over value at 52 hours
    \begin{enumerate}
        \item Windowed value at 57 hours over windowed value at 52 hours
    \end{enumerate}
    \item Ratio of ``value at 57 hours minus value at 52 hours" to ``value at 7 hours minus value at 2 hours"
    \begin{enumerate}
        \item Ratio of ``windowed value at 57 hours minus windowed value at 52 hours" to ``windowed value at 7 hours minus windowed value at 2 hours"
    \end{enumerate}
    \item Coefficients from ARIMA models fit to post-confluence/pre-wounding data, from 24 to 36 hours:
    \begin{enumerate}
        \item $ARIMA(1,1,0)$ ($AR(1)$ model with first differencing), $\hat{\phi}_1$
        \item $ARIMA(1,1,0)$, constant term
        \item $ARIMA(0,1,1)$ ($MA(1)$ model with first differencing), $\hat{\theta}_1$
        \item $ARIMA(0,1,1)$, constant term
    \end{enumerate}
    \item Estimated error variance from these fitted ARIMA models:
    \begin{enumerate}
        \item $ARIMA(1,1,0)$, $\hat{\sigma}^2$
        \item $ARIMA(0,1,1)$, $\hat{\sigma}^2$
    \end{enumerate}
    \item Features measuring autocorrelation time series from 24 to 36 hours, after first differencing:
    \begin{enumerate}
        \item ACF at lag 1
        \item ACF at lag 2
        \item PACF at lag 1
        \item PACF at lag 2
    \end{enumerate}
    \item Sample variance of once-differenced time series from 24 to 36 hours
    \item Sample standard deviation of once-differenced time series from 24 to 36 hours
    \item Features of the first peak in a smoothed time series after wounding:
    \begin{enumerate}
        \item Value at this peak
        \item Time when this peak occurs
    \end{enumerate}
    \item Minimum value after wounding
    \begin{enumerate}
        \item Minimum value after wounding of a smoothed time series
    \end{enumerate}
        \item Features from the Growth-to-Confluence Detector of Zhang et al., applied to post-wounding data:
    \begin{enumerate}
        \item $\tau$, the detected time for the beginning of the confluent phase
        \item $d$, the estimated long-memory parameter for the series following confluence
    \end{enumerate}
\end{enumerate}

\section{Additional results}

\subsection{Results for classification algorithms}

The three classification algorithms used in this study are classification trees, linear discriminant analysis (LDA), and quadratic discriminant analysis (QDA). Following the findings of \cite{gelsinger}, we do not investigate further possibilities for the parameters of the discriminant analysis beyond LDA and QDA, since these were not found to have a great effect on the classification of cells.

Overall, LDA was the most effective classification algorithm on our main dataset. The average performance of the LDA classifiers over all feature pairs was 0.882, as compared to 0.855 for QDA classifiers and 0.857 for trees. In a small number of cases, the discriminant analysis algorithms are unable to define classification regions, thanks to a small sample size and small numerical differences between individual wells' feature scores. The tree method can be used in these cases.

The difference in performance between the methods is, indeed, not large. We find LDA to be desirable, however, not only for its slight advantage in performance but for the nature of its classification regions, which are easy to describe and visualize.

\subsection{Results for different cell/contaminant types}

We summarize additional results from the within-plate classification task for the three different combinations of cell type and contaminant.

For MDCK-II cells with \emph{M. hominis}, 27\% of all feature pairs yielded perfect classification accuracy on BSA-treated cultures, and 25\% of all pairs did so on gel-treated cultures. Some examples of pairs with perfect accuracy include:

\begin{itemize}
    \item     57 hours - 52 hours, 32000 Hz and Value of initial peak, 1000 Hz; BSA
    \item Sample variance of differenced series, 8000 Hz and Value of post-wound peak, 2000 Hz; BSA
    \item Value at 2 hours, 500 Hz and PACF of differenced series, lag 2, 32000 Hz; gel
\end{itemize}

There are many individual features with average accuracy rates extremely close to 1 (above 0.99). These include:

\begin{itemize}
    \item (57 hours - 52 hours)/(7 hours - 2 hours),500 Hz, gel
    \item 57 hours - 52 hours, 500 Hz, BSA
    \item Quadratic coefficient from fit to first 2 hours, 100 Hz, gel
    \item Value of post-wound peak, 1000 Hz, gel
\end{itemize}

For MDCK-II cells with \emph{M. hyorhinis}, 13\% of all feature pairs yielded perfect classification accuracy on BSA-treated cultures, and 38\% of all pairs did so on gel-treated cultures. Some examples of pairs with perfect accuracy include:

\begin{itemize}
    \item Quadratic coefficient from fit to first 2 hours, 16000 Hz and 57 hours/52 hours, 32000 Hz, gel
    \item Max over first 24 hours, 4000 Hz and Quadratic coefficient from fit to first 2 hours, 1000 Hz, gel
    \item Minimum value after wounding, 500 Hz and Post-wounding long-memory parameter $d$, 1000 Hz, BSA
\end{itemize}

Again, many individual features have average accuracy rates above 0.99. These include:

\begin{itemize}
    \item Max over first 24 hours, 8000 Hz, BSA
    \item 57 hours - 52 hours, 32000 Hz, gel
    \item Quadratic coefficient from fit to first 2 hours, 400 Hz, gel
    \item Minimum value after wounding, 500 Hz, gel
\end{itemize}

For BSC-1 cells with \emph{M. hominis}, 0.08\% (888/113469) of all feature pairs yielded perfect classification accuracy on BSA-treated cultures, and only 4 pairs did so on gel-treated cultures. Some examples of pairs with perfect accuracy include:

\begin{itemize}
    \item Max over first 24 hours, 32000 Hz and Max over first 24 hours, 8000 Hz, gel
    \item Sample variance of differenced series, 500 Hz and Value at 24 hours, 500 Hz, BSA
    \item Quadratic coefficient from fit to first 2 hours, 1000 Hz and 7 hours - 2 hours, 16000 Hz, BSA
\end{itemize}

Overall, the individual features with the highest average accuracy rate are:

\begin{itemize}
    \item Quadratic coefficient from fit to first 2 hours, 500 Hz, BSA: 0.964
    \item Sample standard deviation of differenced series, 32000 Hz, BSA: 0.957
    \item Sample standard deviation of differenced series, 1600 Hz, BSA: 0.950
    \item Linear coefficient from quadratic fit to first 2 hours, 1000 Hz, BSA: 0.943
\end{itemize}

\subsection{Results for cross-plate classification}

For the cross-plate classification task, few feature pairs had perfect average classification accuracy across train-test combinations: 1 for BSA-treated cells and 97 for gel-treated cells. Some examples with accuracy above 0.99 include:

\begin{itemize}
    \item $\hat{\sigma}^2$ from ARIMA(1,1,0), 500 Hz and Value of post-wound peak, 500 Hz, BSA 
    \item 7 hours - 2 hours, 4000 Hz and Constant from ARIMA(0,1,1), 16000 Hz, gel
    \item Sample standard deviation of differenced series, 500 Hz and Value at 2 hours, 1000 Hz, gel
\end{itemize}

Overall, the individual features with the highest average accuracy rate are:

\begin{itemize}
    \item 7 hours - 2 hours, 4000 Hz, gel: 0.933
    \item Sample standard deviation of differenced series, 500 Hz, BSA: 0.926
    \item 7 hours - 2 hours, 2000 Hz, gel: 0.925
    \item $\hat{\sigma}^2$ from ARIMA(0,1,1) model, 5000 Hz, BSA: 0.924
\end{itemize}

\bibliographystyle{plain}
\bibliography{abpcontambib}